\documentclass[aps,pre,twocolumn,showpacs,superscriptaddress,groupedaddress]{revtex4}
\usepackage{graphicx}
\usepackage{bm}        
\usepackage{amssymb}   
\begin{document}

\title{Directed translocation of a flexible polymer through a cone-shaped channel }

\author{Narges Nikoofard}
\affiliation{Department of Physics, Institute for Advanced Studies in Basic Sciences (IASBS), Zanjan 45137-66731, Iran}

\author{Hamidreza Khalilian}
\affiliation{Department of Physics, Institute for Advanced Studies in Basic Sciences (IASBS), Zanjan 45137-66731, Iran}

\author{Hossein Fazli}
\email{fazli@iasbs.ac.ir}
\affiliation{Department of Physics, Institute for Advanced Studies in Basic Sciences (IASBS), Zanjan 45137-66731, Iran}
\affiliation{Department of Biological Sciences, Institute for Advanced Studies in Basic Sciences (IASBS), Zanjan 45137-66731, Iran}
\date{\today}

\begin{abstract}
Entropy-driven directed translocation of a flexible polymer through a cone-shaped channel is studied theoretically and using computer simulation. For a given length of the channel, the effective force of entropic origin acting on the polymer is calculated as a function of the apex angle of the channel. It is found that the translocation time is a non-monotonic function of the apex angle. By increasing the apex angle from zero, the translocation time shows a  minimum and then a  maximum. Also, it is found that regardless of the value of the apex angle, the translocation time is a uniformly decreasing function of the channel length. The results of the theory and the simulation are in good qualitative agreement.

\end{abstract}

\maketitle

\section{Introduction}

It is known that the Brownian motion of particles in the presence of asymmetric structures and in non-equilibrium conditions may result in directed motion. This phenomenon, called Brownian motor or Brownian ratchet is ubiquitous in the living cells. The known examples of this phenomenon are the electric potential difference through the ion channels and the movement of the kinesin motor protein along the microtubule  \cite{astumian2001,hanggi2009}. This phenomenon has attracted great interest in recent years, due to its applications in the separation of particles \cite{mahmud2009,matthias2003} and making pumps \cite{siwy2002} and motors \cite{sokolov2010,leonardo2010} in fine dimensions. The directed motion of particles arisen by an asymmetric structure as a non-equilibrium phenomenon has been observed in systems covering a broad range of scales such as macroscopic elastic discs \cite{shaw2007}, mesoscopic gears \cite{sokolov2010,leonardo2010}, microscopic colloidal systems \cite{matthias2003}, moving cells \cite{mahmud2009} and ions \cite{siwy2002}.

In the case of polymers, directed translocation of a polymer through a curved bilayer membrane \cite{baumgartner1995} and  polymer passage through a membrane in the presence of chaperons \cite{hanggi2009,simon1992} have been studied. Despite extended studies in this field, the polymer motion through asymmetric structures such as a cone-shaped channel is poorly understood. It has been shown that cone-shaped channels  have important applications in rectifying the ionic currents and in the simulation of biological ion channels \cite{siwy2002}. Polymer translocation through cone-shaped channels has also been studied, in the literature \cite{sexton2007}. It has been observed that the dependencies of the translocation time and the capture rate on the applied voltage and the polymer length are qualitatively similar to those of the cylindrical channels. The most important point in the case of cone-shaped channels is the very high strength of the electric field in the narrow entry of the channel. It causes the ionic current through the channel to be affected only by the few monomers in the narrow entry of the channel \cite{sexton2007}. This point is important in the application of these channels in DNA sequencing. One important challenge of DNA sequencing by the ordinary channels such as $\alpha$-hemolysin protein channel is the simultaneous effect of 10-15 nucleic acids inside the channel on the ionic current \cite{branton2008}. However, in the cone-shaped protein channel, MspA, only two nucleic acids which are close to the channel apex simultaneously affect the current \cite{derrington2010}. In DNA translocation experiments through MspA, the difference between the effects of the four types of the nucleic acids on the ionic current is larger, compared to the experiments using the cylindrical channels \cite{derrington2010}. This is also an important advantage of the cone-shaped channels  for DNA sequencing \cite{venkatesan2011}.

In this paper, the translocation of a flexible polymer through a cone-shaped channel in the presence of no external driving field is studied, theoretically and by computer simulation.
During the translocation process, a force of entropic origin acts on the part of the polymer which is inside the channel. This force originates from the entropic tendency of the polymer toward the larger entry of the channel. The translocation time is a decreasing function of this force. We set out to obtain  the effective force, $f$, acting on the polymer as a function of the channel apex angle and the channel length and compare it with the simulation results.
For a given length of the cone-shaped channel, we calculate the effective force for the two cases of small and large apex angles of the channel, theoretically. During the translocation of a polymer through a narrow channel in a wall (from the cis to the trans side), it is known that the monomers of the polymer segment passed to the trans side accumulate near the wall, in front of the channel exit. Therefore, to calculate the effective force acting on a polymer which translocates through a cone-shaped channel of small base diameter (from the apex to the base), we consider the channel as a closed cavity confining a segment of the polymer. In the case of large base diameters, however, we consider one of the polymer ends fixed in the apex of the cone-shaped channel and obtain the confinement free-energy and then the exerted force to the polymer.
As a further check of the reliability of the results, the force is calculated from two different methods in the case of large diameters of the channel base.

For a given length of the channel, we find that the effective force  is a non-monotonic function of the channel apex angle. Combination of the results obtained from the two cases of small and large apex angles shows that the force as a function of the apex angle has a maximum and then a minimum. We also obtain the dependence of the force on the channel length. For each value of the channel apex angle, the force is a monotonically increasing function of the channel length. These results are supported by our simulation data for the translocation time of a polymer through a cone-shaped channel. The theory also predicts the polymer length inside the channel, which is in agreement with the simulation results.

The rest of the paper is organized as follows. In Sec. \ref{theory}, we present the theory, for the two cases of small and large apex angles. The force on the polymer and the average number of the monomers inside the channel are discussed. In Sec. \ref{simulation}, the simulation results of the polymer translocation through a cone-shaped channel are presented and compared with the theory. Finally, in the last section, the paper is summarized, the results are discussed and some notes are presented on the polymer equilibrium during its translocation through the cone-shaped channel.

\begin{figure}
\includegraphics[scale=0.55]{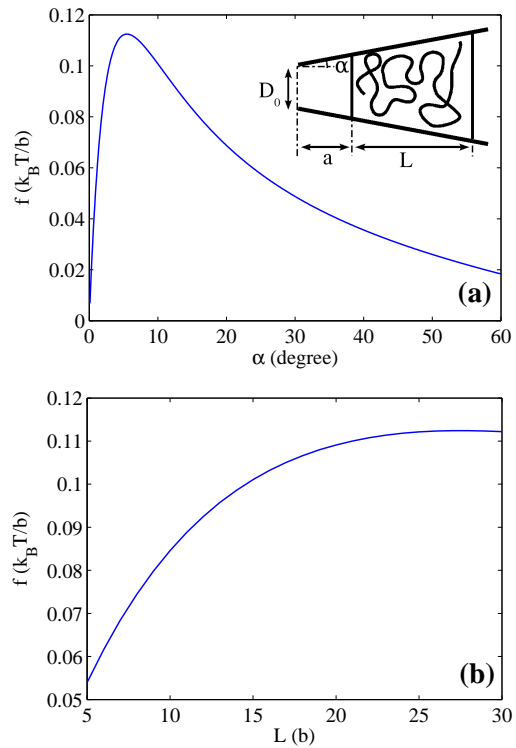}
\caption{(Color online) (a) The effective force exerted to a polymer confined in a closed frustum of length $L=15b$, as a function of the cone apex angle. As can be seen, the effective force has a maximum at small apex angles. Inset: Schematic of a polymer confined in a frustum by two walls perpendicular to the channel axis. (b) The effective force as a function of the frustum length, for a channel of apex angle $\alpha = 3\,^{\circ}$. For a given value of the apex angle, the effective force monotonically increases with  the frustum length. $D_0=2b$ for both panels (a) and (b).  } \label{closed-force}
\end{figure}

\section{Theory}    \label{theory}

Simulation results (in the next section) show that the polymer translocation through a cone-shaped channel is a driven process. When one end of a polymer is fixed in the channel apex, a driving force is exerted on the polymer, by the channel. This force results from the increase in the polymer entropy with moving to the wider parts of the channel. During the polymer translocation, also, the very narrow apex of the channel divides the polymer into separate parts with separate free energies. The free energy of the polymer part inside the channel decreases with the polymer movement toward the wide entry of the channel. Indeed, the system is in nonequilibrium and writing a total free energy for the polymer is meaningless. Hence, in this section, we consider the part of the polymer which is inside the channel as a separate polymer in a conical channel and calculate its free energy. The derivative of the free energy gives the force exerted on the polymer, which is the driving force in the polymer translocation. We discuss the assumption of polymer in equilibrium in Sec. \ref{discuss}.

It has already been shown that during the polymer translocation through a narrow channel, the monomers of the polymer segment passed to the opposite side crowd close to the channel \cite{kantor2004,bhattacharya2010}. Accordingly, for the case of small apex angles of the channel, we assume that the channel is a closed volume containing a part of the polymer. With this assumption, we calculate the force acting on the polymer due to the asymmetric shape of the channel. In the case of large apex angles, we calculate the force acting on a polymer whose end is fixed in the apex of a cone-shaped channel and calculate the driving force on the polymer, using two different methods. By combining the results of the two cases, we conclude the qualitative behavior of the force acting on the polymer as a function of the channel apex angle.

\subsection{The case of small apex angles}

Consider a flexible polymer, consisting of $N$ spherical monomers of diameter $b$, which is confined inside a frustum (see the inset of Fig. \ref{closed-force}(a)). The confinement free energy of a polymer inside a closed space of volume $\Omega$ is known to be $F \sim k_BT\left(\frac{N^{3\nu}b^3}{\Omega}\right)^{\frac{1}{3\nu-1}}$, where $\nu$ is the Flory exponent \cite{sakaue2006}. In our case, the volume of the confining space is equal to
$\Omega \sim \frac{1}{\tan\alpha}\left((D_0+2(a+L)\tan\alpha)^3-(D_0+2a\tan\alpha)^3\right)$.
Therefore, the free energy of confinement can be written as
\begin{widetext}
\begin{equation} \label{small-energy}
F \sim k_BT \left(\frac{N^{3\nu}b^3\tan\alpha}{(D_0+2(a+L)\tan\alpha)^3-(D_0+2a\tan\alpha)^3}\right)^{1/(3\nu-1)}.
\end{equation}
\end{widetext}
Here, $\alpha$ is half of the apex angle of the channel, and $a$ is the distance between the region in which the polymer is confined and a cross section of the cone with diameter $D_0$. $L$ is the length of the region that confines the polymer. When the confinement region is moved toward the wider parts of the channel (increasing $a$ in the inset of Fig. \ref{closed-force}(a)), the entropy of the polymer increases and its confinement free-energy decreases. In other words, if one removes the two confining walls, which are perpendicular to the channel axis, the polymer moves to the wider part of the channel to gain more entropy. This is the origin of the force exerted to the segment of a polymer inside the channel, in the course of its translocation through a cone-shaped channel. To obtain the driving force in the polymer translocation through a cone-shaped channel of length $L$ and tip diameter $D_0$, we calculate the derivative of the confinement free-energy (Eq. \ref{small-energy}) with respect to $a$, at $a=0$; $f = -\left(\frac{\partial F}{\partial a}\right)_{a=0}$.
One should note here that for a translocating polymer, $N$ in Eq. \ref{small-energy} is the number of monomers inside the channel, not the total length of the polymer.
The force, $f$, acting on the polymer segment inside the channel is obtained as
\begin{widetext}
\begin{equation} \label{small-force}
f \sim \frac{k_BT}{D_0} \left(\frac{N^{\nu}b}{D_0}\right)^{\frac{3}{3\nu-1}} \left(\frac{\tan\alpha}{\left(1+2\frac{L}{D_0}\tan\alpha\right)^3-1}\right)^{\frac{3\nu}{3\nu-1}} \left(\left(1+2\frac{L}{D_0}\tan\alpha\right)^2-1\right).
\end{equation}
\end{widetext}
In this equation, the number of monomers inside the channel, $N$, should be substituted from Eq. \ref{N} (see below). In Figs. \ref{closed-force}(a) and \ref{closed-force}(b), the force, $f$, is shown as a function of the channel apex angle, $\alpha$, for a given length of the channel, and as a function of the channel length, $L$, for a given value of $\alpha$.  As can be seen in Fig. \ref{closed-force}(a), the force as a function of the apex angle has a maximum at small values of $\alpha$. The existence of this maximum can be explained regarding the two determining factors in the force: the strength of the polymer confinement in the channel and the magnitude of the asymmetry in the channel shape. At small apex angles, the channel volume is small and the strength of the confinement  is high. By increasing the apex angle and hence the asymmetry of the channel shape, the force of entropic origin exerted to the polymer increases. By more increasing the apex angle, the confinement effect of the channel on the polymer weakens and the strength of the force decreases. The maximum value of the force corresponds to a value of the apex angle for which the combination of the confinement effect and  the asymmetric shape of the channel has the optimum driving effect.

\subsection{The case of large apex angles}
\begin{figure}
\includegraphics[scale=0.35]{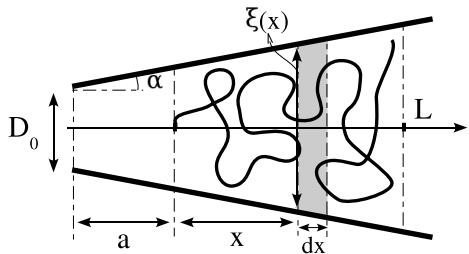}
\caption{Schematic of a polymer inside a long cone-shaped channel. One end of the polymer is considered to be fixed in a cross section of the channel. The blob diameter as defined in the first method of the theory (in the case of large apex angles) is shown in the figure.} \label{open-conic}
\end{figure}

In the case of large apex angles, considering the polymer as a confined one in a closed volume is not reasonable. Instead, we consider a polymer that one of its ends is fixed in a cross section of a long cone-shaped channel and use the blob method to calculate the effective driving force. For a conical channel, the diameter of a blob depends on its position along the channel. It is equal to the channel diameter at each position; $\xi(x)=D_0+2(a+x)\tan\alpha$ (see Fig. \ref{open-conic}). Here, $a$ is the distance of the fixed end of the polymer from a cross section of the channel with diameter $D_0$.
Following Ref. \cite{werner2010}, we write the confinement free energy of the polymer as $\frac{F}{k_BT} \sim \int_0^L \frac{dx}{\xi(x)}$, which gives
\begin{widetext}
 \begin{equation} \label{large-energy}
\frac{F}{k_BT} \sim \frac{1}{2\tan\alpha} \left[\ln\left(D_0+2(a+L)\tan\alpha\right)-\ln\left(D_0+2a\tan\alpha\right)\right].
\end{equation}
\end{widetext}
Here, $L$ is  the polymer size along the channel.
The number of monomers inside each blob is $g(x) \sim \left(\frac{\xi(x)}{b}\right)^{\frac{1}{\nu}}$ \cite{rubinstein2003}. Also, the number density of the monomers inside each blob is $\rho(x)\sim\frac{g(x)}{\xi(x)^3}$, and the cross-section area of each blob scales as $\xi(x)^2$. Hence, the number of monomers inside a region of thickness $dx$ inside the channel (the region colored in gray in Fig. \ref{open-conic}) can be written as $n(x)dx \sim \frac{g(x)}{\xi(x)} dx$ \cite{werner2010}.
The size of the polymer along the channel can be obtained by the equality of the integral of $n(x)$ over $x$ and the total number of the monomers, $N$;
\begin{widetext}
\begin{equation} \label{N}
N \sim \int_0^L \frac{g(x)}{\xi(x)} dx = \frac{b^{-\frac{1}{\nu}}}{\tan\alpha} \left[\left(D_0+2(a+L)\tan\alpha\right)^{\frac{1}{\nu}}-\left(D_0+2a\tan\alpha\right)^{\frac{1}{\nu}}\right].
\end{equation}
\end{widetext}
The force exerted to the polymer can be calculated from the derivative of the confinement free energy, Eq. \ref{large-energy}, with respect to $a$, at $a=0$. Here, one should note that $N$ is the total number of the monomers on the polymer and its value is constant. Instead, the polymer size along the channel axis, $L$, depends on the parameters such as $a$ (Eq. \ref{N}).
From Eq. \ref{large-energy}, we can calculate the force $f$;
$\frac{f}{k_BT} \sim -\left(1+\left(\frac{\partial L}{\partial a}\right)_{a=0}\right)\frac{1}{D_0+2L\tan\alpha}+\frac{1}{D_0}$.
The derivative of $L$ with respect to $a$ is found from Eq. \ref{N},
$\left(\frac{\partial L}{\partial a}\right)_{a=0} \sim (\frac{D_0}{D_0+2L\tan\alpha})^{\frac{1}{\nu}-1}-1$.
Accordingly,
\begin{equation} \label{large-force}
\frac{f}{k_BT} \sim \frac{1}{D_0}\left[1-\left(\frac{D_0}{D_0+2L\tan\alpha}\right)^{\frac{1}{\nu}}\right].
\end{equation}
To calculate the force in the polymer translocation through a cone-shaped channel, $L$ in Eq. \ref{large-force} should be substituted by the channel length. The assumption of constant $N$ and variable $L$ in the calculation of the force is also valid for the case of polymer translocation. Because, at each moment, the polymer segment inside the channel does not feel the finite length of the channel.

\begin{figure}
\includegraphics[scale=0.4]{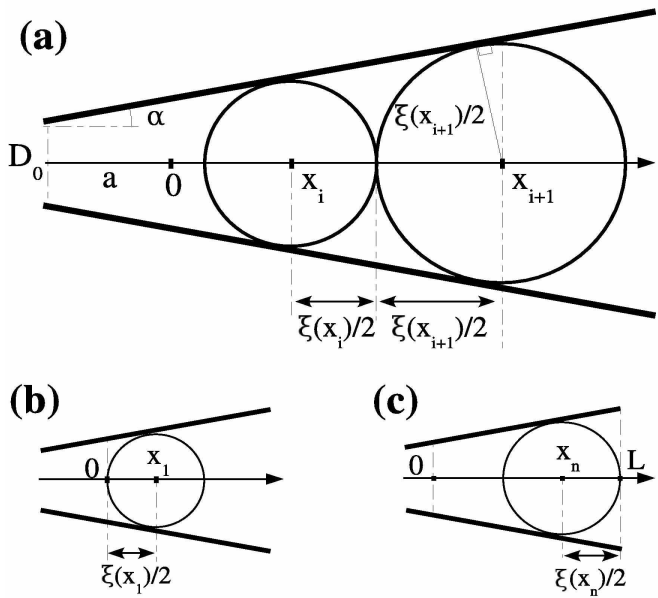}
\caption{(a) Definition of the blob in the second method of the theory (the case of large apex angles). The size of the blob at position $x$ on the channel axis, and the relation between the positions of two consecutive blobs are shown in the figure. (b), (c) The first blob is tangent to the beginning of the channel, and the last blob is tangent to its end.} \label{open-conic2}
\end{figure}

The exerted force can also be calculated in another way, whose algebraic calculations are more complicated. This re-calculation is useful for checking the results.
For this end, the blobs are defined as spheres tangent to the channel wall that cannot penetrate into each other. The same method for defining the blobs has been used in Ref. \cite{brochard1993}. The blobs are assumed as spheres tangent to the cone, so, the radius of a blob with its center at position $x$ is $\xi(x) = D_0\cos\alpha + 2(x+a)\sin\alpha$ (see Fig. \ref{open-conic2}(a)). Confinement free energy of the polymer is proportional to the number of these blobs. To count the blobs and calculate the confinement free energy, we use the geometrical relation between the positions of two consecutive blobs inside the channel;
$x_{i+1} - x_i = \frac{\xi(x_i)+\xi(x_{i+1})}{2}$ (see Fig. \ref{open-conic2}(a)).
Using this recursive relation, one can find the explicit relation between the position of a blob, $x_i$, and its number along the channel, $i$;
$x_i = A^{i-1} \left(x_1+a+\frac{D_0}{2}\cot\alpha\right) - \left(a+\frac{D_0}{2}\cot\alpha\right)$,
where $A = \frac{1+\sin\alpha}{1-\sin\alpha}$.
The first and the last blobs are tangent to  the beginning and the end of the channel, respectively. Hence, their positions along the channel, $x_1$ and $x_n$, can be obtained from the relations $x_1 = \frac{\xi(x_1)}{2}$ and $x_n+\frac{\xi(x_n)}{2} = L$, respectively. $n$ is the total number of the blobs (see Fig. \ref{open-conic2}).
Using the equations for $x_i$, $x_1$ and $x_n$, the number of the blobs, $n$, and the confinement free energy are obtained as
\begin{widetext}
\begin{equation}  \label{large-energy2}
\frac{F}{k_BT} \sim n \sim \frac{\log\left[2(L+a)\sin\alpha +D_0\cos\alpha\right]-\log\left[2a\sin\alpha +D_0\cos\alpha\right]}{\log(1+\sin\alpha)-\log(1-\sin\alpha)}.
\end{equation}
\end{widetext}
The polymer extension along the channel axis can be obtained from the constraint $N = \sum_{i=1}^{n} g(x_i)$, where $g(x_i)$ is the number of monomers inside the $i$th blob. Substituting $g(x_i) \sim (\frac{\xi(x_i)}{b})^{\frac{1}{\nu}}$ and $\xi(x_i) \sim D_0\cos\alpha + 2(x_i+a)\sin\alpha$, and using the equation for $x_i$, one obtains
\begin{widetext}
\begin{equation} \label{N2}
N = b^{-\frac{1}{\nu}}\frac{\left[2(L+a)\sin\alpha +D_0\cos\alpha\right]^{\frac{1}{\nu}}-\left[2a\sin\alpha +D_0\cos\alpha\right]^{\frac{1}{\nu}}}{(1+\sin\alpha)^{\frac{1}{\nu}}-(1-\sin\alpha)^{\frac{1}{\nu}}}.
\end{equation}
\end{widetext}
Using the derivatives of Eqs. \ref{large-energy2} and \ref{N2}, we have
\begin{equation} \label{large-force2}
\frac{f}{k_BT} \sim \frac{1}{D_0} \frac{\tan\alpha}{\log\left(\frac{1+\sin\alpha}{1-\sin\alpha}\right)}\left[1-\left(\frac{D_0}{D_0+2L\tan\alpha}\right)^{\frac{1}{\nu}}\right].
\end{equation}
Note that Eq. \ref{large-force2} is quite similar to the result obtained in Eq. \ref{large-force}, and differs only on the coefficient $\frac{\tan\alpha}{\log\left(\frac{1+\sin\alpha}{1-\sin\alpha}\right)}$.

\begin{figure}
\includegraphics[scale=0.55]{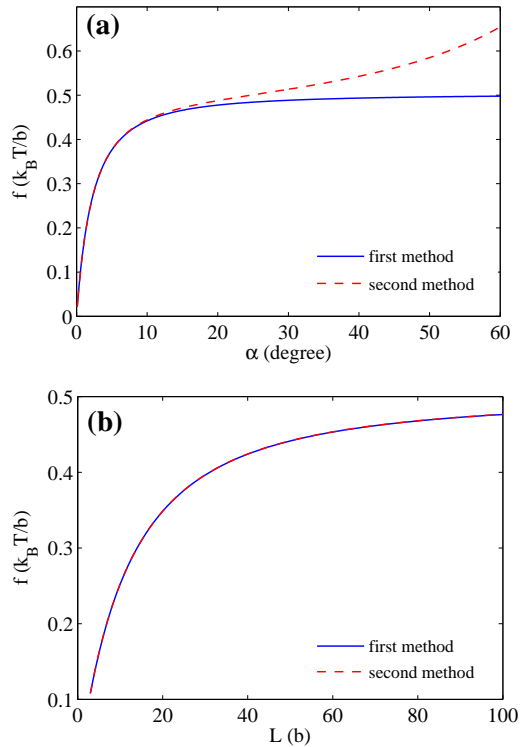}
\caption{(Color online) The force exerted to a polymer, which is confined inside an open cone-shaped channel, versus (a) the cone apex angle, with the channel length $L=15b$, and (b) the channel length, with the apex angle $\alpha=3\,^{\circ}$. In these figures, $D_0=2b$. The force is calculated from the two methods described in the text. When the apex angle is small, the two methods have the same results. The force increases with the cone angle and length and becomes constant afterwards.} \label{open-force}
\end{figure}

When the diameter of the first blob is smaller than the length of the channel, $\xi(x_1)<L$, we expect the results obtained from this method to be the same as those of the previous method. The value of $\xi(x_1)$ does not depend on the channel length, but it increases with the apex angle, rapidly. As is shown in Fig. \ref{open-force}, the two methods give the same dependence of the force on the channel length. Also, dependence of the force on the apex angle calculated from the two methods is the same at small apex angles, and becomes different only at large angles.
In the case of large apex angles, despite the case of small ones, the force increases monotonically with  the apex angle (Fig. \ref{open-force}(a)). The force increases with the channel length, $L$, and becomes constant, at larger values of $L$ (Fig. \ref{open-force}(b)). Indeed, at large values of $L$, the channel diameter becomes larger than the radius of gyration of the polymer and it does not have any confining effect on the polymer.

\begin{figure}
\includegraphics[scale=0.55]{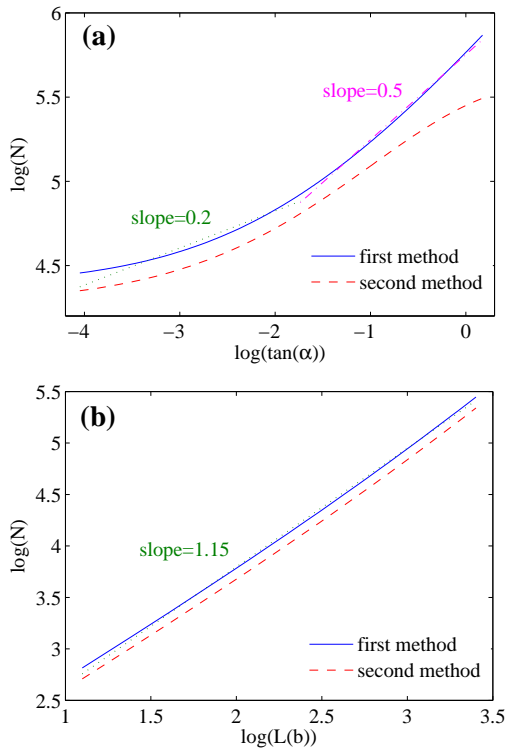}
\caption{(Color online) The length (number of monomers) of the polymer segment inside the channel versus (a) the channel apex angle (for $L=15b$), and (b) the length of the channel (for $\alpha=3\,^{\circ}$). To make the comparison easier, the polymer length obtained from the second method is multiplied by 3, in both panels (a) and (b). As can be seen, the two methods predict the same dependence of the polymer length on the channel length and the apex angle. Power-law fits to the curves are shown, for later comparison with the simulation results (see Sec. \ref{simulation}).} \label{blobs-N}
\end{figure}

The length of the polymer segment inside the channel, $N$, versus the channel length and the apex angle is shown in Figs. \ref{blobs-N}(a) and \ref{blobs-N}(b). Although the values of $N$ obtained from the two methods  are different, they have the same dependence on the channel length and the apex angle. For comparison with the simulation results presented in the next section, power-law functions are fitted to the results. In Fig. \ref{blobs-N}(a), two power-law functions with exponents $0.20$ and $0.50$ are fitted to the result of the theory, at small and large apex angles, respectively. Also, as is shown in Fig. \ref{blobs-N}(b), the curve obtained from the theory is followed well by a power-law function with exponent $1.15$.

\section{simulation} \label{simulation}

\subsection{The simulation method}
We use molecular dynamics (MD) simulations to check our theory. We use the coarse-grained bead-spring model for the polymer. The interaction between the monomers is the short-ranged Lennard-Jones repulsive potential
\begin{equation}
 U_{LJ}(r) = \left\lbrace
  \begin{array}{l l}
    4\varepsilon\
    \{(\frac{\sigma}{r})^{12}-(\frac{\sigma}{r})^{6}+\frac{1}{4}\} \qquad & r < 2^{1/6}\sigma,\\
    0 \qquad & r \geq 2^{1/6}\sigma,
  \end{array}
\right.
\label{lj}
\end{equation}
where $\sigma$ is the monomer diameter and the MD length scale, and $\varepsilon$ is the Lennard-Jones energy scale. The monomers are connected by the harmonic potential, $U_{harm}=-\frac{1}{2}K(r-R)^2$, in which $K$ is the spring constant, $r$ is the distance between two consecutive monomers along the polymer and $R$ is their equilibrium distance.
The cone-shaped channel and the two walls are modeled by the Lennard-Jones potential between them and the monomers (see the schematic of the channel and the polymer in the inset of Fig. \ref{conic-biased}).

The time step of the simulations is $\tau=0.01\tau_0$, where $\tau_0=\sqrt{\frac{m\sigma^2}{\varepsilon}}$ is the MD time scale, and $m$ is the monomer mass. The simulations are performed in the NVT ensemble, using the Langevin thermostat, at the constant temperature $T=1\varepsilon/k_B$. The Langevin equation $m\ddot{\vec{r}}=-\vec{\nabla} U(\vec{r})-\xi_b \dot{\vec{r}}+\vec{F}_{ext}+\vec{\eta}(t)$ is integrated, for describing the monomers motion, where $\xi_b$ is the friction coefficient, and $\vec{F}_{ext}$ is the external force from the channel and the walls.  $\vec{\eta}(t)$ is the Gaussian white noise, which follows the fluctuation-dissipation theorem. The simulations are done with ESPResSo \cite{espresso}.

At the beginning of the simulation, the monomers are arranged on the axis of the channel, such that the lengths of the two segments of the polymer, which are outside the channel from the two sides, are the same. Then, we fix the monomer in the channel apex and let the other monomers to equilibrate. After that, we release the fixed monomer and set the simulation time equal to zero. The passage time is the time that the last monomer of the polymer lies in the channel apex.  In our simulations, the diameter of the channel apex is $D_0=2.4b$ and the polymer has 100 monomers, unless otherwise stated.

\subsection{Simulation results}
\begin{figure}
\includegraphics[scale=0.5]{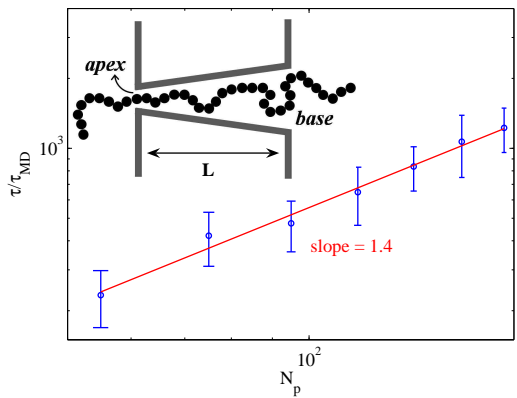}
\caption{(Color online) The translocation time of a polymer through a cone-shaped channel, versus the polymer length. The channel parameters are $L=15b$ and $\alpha=3\,^{\circ}$. The exponent is close to the case of the driven polymer translocation \cite{ikonen2012}. The error-bars are resulted from averaging over 21 runs of the simulation. Inset: schematic of the cone-shaped channel, the two walls and the polymer used in the simulation.} \label{conic-biased}
\end{figure}

When the channel is cylindrical ($\alpha=0$), the polymer exits the channel with the same probabilities from the two sides of the channel. The polymer passage time in this case scales with the polymer length by the exponent 2.2. This is consistent with the previous results for unbiased polymer translocation \cite{luo2008}. Instead, in the translocation from the cone-shaped channel, the polymer leaves the channel from the base in all our simulations and the passage time scales with the polymer length with the exponent 1.4 (see Fig. \ref{conic-biased}). This exponent is close to the exponent 1.6 predicted in the literature, for the driven translocation (of infinitely long polymers) \cite{kantor2004, ikonen2012}. It also completely matches the exponent for the driven translocation of shorter polymers \cite{luo2008}. This shows that the force exerted from the cone-shaped channel is determining and the polymer translocation through this channel is a driven process.

\begin{figure}
\includegraphics[scale=0.55]{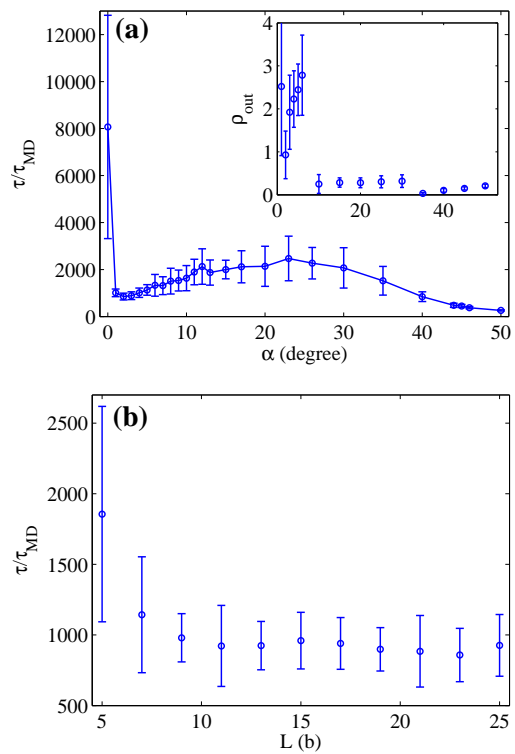}
\caption{(Color online) The simulation results for the passage time of the polymer versus (a) the channel apex angle (for $L=15b$), and (b) the channel length (for $\alpha=3\,^{\circ}$). The passage time is a non-monotonic function of the apex angle, but decreases monotonically with the length. The inset of panel (a) shows the monomers density outside the channel base. It shows that the monomers crowd close to the channel. The error-bars are resulted from averaging over 15 runs of the simulation. For the zero apex angle, 42 runs of the simulation are done.} \label{tau-sim}
\end{figure}

The polymer passage time versus the channel apex angle is shown in Fig. \ref{tau-sim}(a), for $L=15b$. Upon increasing the channel angle, $\alpha$, from $0\,^{\circ}$, the passage time decreases and shows a  minimum at $2\,^{\circ}$. Then the passage time increases up to its maximum value at $25\,^{\circ}$ and then decreases  at larger angles. Note that at small angles, the channel base diameter is small and the simulation data should be described by the case of small apex angles of the theory, Eq. \ref{small-force}. The force obtained from this equation, first increases and then decreases with the apex angle, as shown in Fig. \ref{closed-force}(a). Considering that the passage time is a decreasing function of the force, it would have the same behavior as the simulation result. However, at large apex angles, the channel base diameter is large and Eq. \ref{large-force} should be used to describe the force exerted to the polymer. The force that is obtained from the case of large apex angles of the theory increases monotonically with the angle (Fig. \ref{open-force}(a)). This is in agreement with the passage time reduction at large apex angles. Combination of the predictions of the two cases of small and large apex angles of the theory describes the simulation result, reasonably well.

In the theory, in the case of the small apex angles, we assumed that the channel is a closed volume. To justify this assumption, the average density of the monomers in the range of distance $3\sigma$ from the channel base is calculated in the simulations (see the inset of Fig. \ref{tau-sim}(a)). It can be seen that the monomers density close to the channel base is higher, in the case of small apex angles.  The sudden decrease in the value of the monomers density occurs where the base diameter becomes of the order of the monomer diameter; $L\tan(\alpha)\sim b$.

The polymer passage time versus the channel length is shown in Fig. \ref{tau-sim}(b), for the apex angle $\alpha=3\,^{\circ}$. This time decreases with the channel length and then becomes constant. From the theory, the force in the case of small base diameters increases with the channel length, at small values of $L$ (Eq. \ref{small-force} and Fig. \ref{closed-force}(b)). Also, in the case of large base diameters, the force increases with $L$ monotonically (Eq. \ref{large-force} and Fig. \ref{open-force}(b)). These are in agreement with the result shown in Fig. \ref{tau-sim}(b).

\begin{figure}
\includegraphics[scale=0.55]{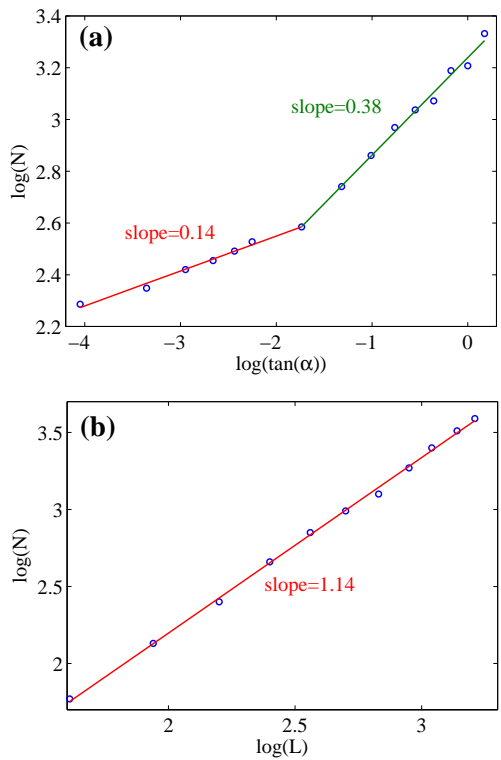}
\caption{(Color online) The number of the monomers of the polymer segment inside the channel  as a function of (a) the channel apex angle (for $L=15b$), and (b) the channel length (for $\alpha=3\,^{\circ}$). As is shown, the fit exponents are close to those of Fig. \ref{blobs-N}.} \label{Nin}
\end{figure}

It is worth studying the number of monomers of the polymer segment inside the channel during its passage through the channel. The mean value of the monomers inside the channel as a function of the channel angle and length is shown in Figs. \ref{Nin}(a) and \ref{Nin}(b). The number of monomers versus the channel length can be  described by a power-law function with exponent $1.14$. However, the number of monomers against the tangent of the channel apex angle can be fitted with two different exponents $0.13$ and $0.38$, for the small and large apex angles, respectively. These figures are plotted for the same parameters as those of Fig. \ref{Nin} and the fit exponents are nearly close.

\section{Summary and discussion}\label{discuss}
An important assumption in our theoretical calculation of the entropic force exerted to the polymer is that the polymer is in equilibrium inside the channel. However, it is known that the polymer passage through a channel is a non-equilibrium process, and it has been shown that the two ends of the polymer out of the channel cannot equilibrate during the passage process \cite{kantor2004,bhattacharya2010}.
Here, we investigate the validity of the assumption that the polymer segment inside the channel is in equilibrium during the passage process. For this end, we compare the relaxation time of the this segment with the time needed for the polymer to traverse the channel length, $L$. If the former time is smaller, the assumption is reasonable.
The velocity of the polymer passing the channel can be written as $v \sim \frac{f}{N_p\xi_b}$, where $f$ is the entropic force exerted to the polymer, $N_p$ the total length of the polymer, and $\xi_b$ the friction constant for each monomer. In this relation, the friction force acting on the polymer is calculated from the Rouse model \cite{rouse1953}. Therefore, the time needed for the polymer to traverse the channel length would be $\tau_L \sim \frac{L}{v} \sim \frac{LN_p\xi_b}{f}$ \cite{kantor2004}.
The relaxation time of the polymer segment inside the channel from the Rouse model is $\tau_{eq} \sim \tau_b N^{1+2\nu}$, where $N$ is the length of the polymer segment inside the channel. Here, one should note that by using such scaling relations, that their numerical pre-factors are not known, judgment on our equilibrium assumption is not possible.
For this reason, we perform simulations to check the polymer equilibrium inside the channel.  In separate simulations, one end of a polymer is kept fixed at the channel apex and after equilibration of the polymer, the density profile of the monomers inside the channel is measured . The density profile of the monomers inside the channel is also obtained during the passage of a long polymer (in the simulations of Sec. \ref{simulation}). Comparison of the results of the two simulations shows the assumption that the polymer segment inside the channel is in equilibrium, is reasonable, in the range of the parameters used in our studies. This shows that in our simulations the polymer length inside the channel is small enough to equilibrate in a short time.

In summary, the effective force of entropic origin acting on a polymer in the course of its passage through a cone-shaped channel was calculated and compared with the results of coarse-grained MD simulations. The force was obtained for the two cases of small and large apex angles of the channel. Combination of the results of the two cases showed that the effective force exerted to the polymer is a non-monotonic function of the channel apex angle. The force increases monotonically with the channel length.
The simulations showed the importance of the force exerted by the channel. It was shown that the simulation results for the polymer passage time through a cone-shaped channel can be described with the theory. Also, it was shown that the simulation and the theoretical results for the polymer length inside the channel during the passage of a long polymer are in good agreement. The simulation results also support the assumption that the polymer segment inside the channel is in equilibrium during the polymer passage.


\begin{thebibliography}{2}

\bibitem{astumian2001}
R. D. Astumian, Scientific American \textbf{285}, 44 (2001).

\bibitem{hanggi2009}
P. Hanggi, and F. Marchesoni, Reviews of Modern Physics \textbf{81}, 387 (2009).

\bibitem{mahmud2009}
G. Mahmud, C. J. Campbell, K. J. M. Bishop, Y. A. Komarova, O. Chaga, S. Soh, S. Huda, K. Kandere-Grzybowska, and B. A. Grzybowski, Nature Physics \textbf{5}, 606 (2009).

\bibitem{matthias2003}
S. Matthias, and F. Muller, Nature \textbf{424}, 53 (2003).

\bibitem{siwy2002}
Z. Siwy, and A. Fulinski, Physical Review Letters \textbf{89}, 198103 (2002).

\bibitem{sokolov2010}
A. Sokolov, M. M. Apodaca, B. A. Grzybowski, and I. S. Aranson, Proc. Natl. Acad. Sci. U.S.A. \textbf{107}, 969 (2010).

\bibitem{leonardo2010}
R. Di Leonardo, L. Angelani, D. Dell’Arciprete, G. Ruocco, V. Iebba, S. Schippa, M. P. Conte, F. Mecarini, F. De Angelis, and E. Di Fabrizio, Proc. Natl. Acad. Sci. U.S.A. \textbf{107}, 9541 (2010).

\bibitem{shaw2007}
R. S. Shaw, N. Packard, M. Schroter, H. L. and Swinney, Proc. Natl. Acad. Sci. U.S.A. \textbf{104}, 9580 (2007).

\bibitem{baumgartner1995}
A. Baumgartner, and J. Skolnick, Physical Review Letters \textbf{74}, 2142 (1995).

\bibitem{simon1992}
S. M. Simon, C. S. Peskin, G. F. and Oster, Proc. Natl. Acad. Sci. U.S.A.  \textbf{89}, 3770 (1992).

\bibitem{sexton2007}
L. T. Sexton, L. P. Horne, and C. R. Martin, Molecular BioSystems \textbf{3}, 667 (2007).

\bibitem{branton2008}
D. Branton \emph{et al.}, Nature Biotechnology \textbf{26}, 1146 (2008).

\bibitem{derrington2010}
I. M. Derrington, T. Z. Butler, M. D. Collins, E. Manrao, M. Pavlenok, M. Niederweis, and J. H. Gundlach, Proc. Natl. Acad. Sci. U.S.A. \textbf{107}, 16060 (2010).

\bibitem{venkatesan2011}
B. M. Venkatesan, and R. Bashir, Nature Nanotech. 6, 615 (2011)

\bibitem{kantor2004}
Y. Kantor, and M. Kardar, Physical Review E \textbf{69}, 021806 (2004).

\bibitem{bhattacharya2010}
A. Bhattacharya, and K. Binder, Physical Review E \textbf{81}, 041804 (2010).

\bibitem{sakaue2006}
T. Sakaue, and E. Raphael, Macromolecules \textbf{39}, 2621 (2006).

\bibitem{werner2010}
M. Werner, and J. U. Sommer, The European Physical Journal E \textbf{31}, 383 (2010).

\bibitem{rubinstein2003}
M. Rubinstein, and R. H. Colby, \emph{Polymer Physics}, Oxford University Press, Oxford (2003).

\bibitem{brochard1993}
F. Brochard-Wyart, Europhysics Letters \textbf{23}, 105 (1993).

\bibitem{espresso}
H. J. Limbach, A. Arnold, B. A. Mann, C. Holm, Comp. Phys. Communications \textbf{174}, 704 (2006).

\bibitem{luo2008}
K. Luo, S. T. T. Ollila, I. Huopaniemi, T. Ala-Nissila, P. Pomorski, M. Karttunen, S.-C. Ying, and A. Bhattacharya, Physical Review E \textbf{78}, 050901(R) (2008).

\bibitem{ikonen2012}
T. Ikonen, A. Bhattacharya, T. Ala-Nissila, and W. Sung, Physical Review E \textbf{85}, 051803 (2012).

\bibitem{rouse1953}
P. E. Rouse, J. Chem. Phys. \textbf{21}, 1272 (1953).

\end{thebibliography}
\end{document}